\documentclass[reprint,superscriptaddress,numerical,showpacs,amsmath,amssymb,aps,prd,floatfix]{revtex4-1}
\usepackage{graphicx}

\usepackage[%
  colorlinks=true,
  urlcolor=blue,
  linkcolor=blue,
  citecolor=blue
]{hyperref}

\usepackage[pdftex]{color}
\usepackage[usenames,dvipsnames,svgnames,table]{xcolor}

\let\oldciteauthor=\citeauthor
\def\citeauthor#1{\hypersetup{citecolor=black}\oldciteauthor{#1}}

\let\oldciten=\onlinecite
\def\onlinecite#1{\hypersetup{citecolor=blue}\oldciten{#1}}

\let\oldcite=\cite
\def\cite#1{\hypersetup{citecolor=blue}\oldcite{#1}}

\makeatletter
\DeclareFontFamily{OMX}{MnSymbolE}{}
\DeclareSymbolFont{MnLargeSymbols}{OMX}{MnSymbolE}{m}{n}
\SetSymbolFont{MnLargeSymbols}{bold}{OMX}{MnSymbolE}{b}{n}
\DeclareFontShape{OMX}{MnSymbolE}{m}{n}{
    <-6>  MnSymbolE5
   <6-7>  MnSymbolE6
   <7-8>  MnSymbolE7
   <8-9>  MnSymbolE8
   <9-10> MnSymbolE9
  <10-12> MnSymbolE10
  <12->   MnSymbolE12
}{}
\DeclareFontShape{OMX}{MnSymbolE}{b}{n}{
    <-6>  MnSymbolE-Bold5
   <6-7>  MnSymbolE-Bold6
   <7-8>  MnSymbolE-Bold7
   <8-9>  MnSymbolE-Bold8
   <9-10> MnSymbolE-Bold9
  <10-12> MnSymbolE-Bold10
  <12->   MnSymbolE-Bold12
}{}

\let\llangle\@undefined
\let\rrangle\@undefined
\DeclareMathDelimiter{\llangle}{\mathopen}%
                     {MnLargeSymbols}{'164}{MnLargeSymbols}{'164}
\DeclareMathDelimiter{\rrangle}{\mathclose}%
                     {MnLargeSymbols}{'171}{MnLargeSymbols}{'171}
\makeatother

\newcommand{\ket}[1]{| #1 \rangle}
\newcommand{\bra}[1]{\langle#1 |}

\begin{document}

\title{Tenfold way and many-body zero modes in the Sachdev-Ye-Kitaev model}

\author{Jan Behrends}
\affiliation{Max-Planck-Institut f{\"u}r Physik komplexer Systeme, 01187 Dresden, Germany}
\affiliation{T.C.M. Group, Cavendish Laboratory, University of Cambridge, J.J. Thomson Avenue, Cambridge, CB3 0HE, United Kingdom}
\author{Jens H. Bardarson}
\affiliation{Department of Physics, KTH Royal Institute of Technology, Stockholm, SE-106 91 Sweden}
\author{Benjamin B\'eri}
\affiliation{T.C.M. Group, Cavendish Laboratory, University of Cambridge, J.J. Thomson Avenue, Cambridge, CB3 0HE, United Kingdom}
\affiliation{DAMTP, University of Cambridge, Wilberforce Road, Cambridge, CB3 0WA, United Kingdom}

\begin{abstract}
The Sachdev-Ye-Kitaev (SYK) model, in its simplest form, describes $k$ Majorana fermions with random all-to-all four-body interactions.
We consider the SYK model in the framework of a many-body Altland-Zirnbauer classification that sees the system as belonging to one of eight (real) symmetry classes depending on the value of $k\mod 8$.
We show that, depending on the symmetry class, the system may support exact many-body zero modes with the symmetries also dictating whether these may have a nonzero contribution to Majorana fermions, i.e., single-particle weight.
These zero modes appear in all but two of the symmetry classes.
When present, they leave clear signatures in physical observables that go beyond the threefold (Wigner-Dyson) possibilities for level spacing statistics studied earlier.
Signatures we discover include a zero-energy peak or hole in the single-particle spectral function, depending on whether symmetries allow or forbid zero modes to have single-particle weight.
The zero modes are also shown to influence the many-body dynamics, where signatures include a nonzero long-time limit for the out-of-time-order correlation function.
Furthermore, we show that the extension of the four-body SYK model by quadratic terms can be interpreted as realizing the remaining two complex symmetry classes;
we thus demonstrate how the entire tenfold Altland-Zirnbauer classification may emerge in the SYK model.
\end{abstract}

\maketitle

\section{Introduction}

The SYK model~\cite{Sachdev:1993hv,Kitaev2015}, named after Sachdev, Ye, and Kitaev, is a zero-dimensional system with random four-body interactions between a set of Majorana fermions. 
It is believed to provide a toy model for the holographic principle~\cite{Maldacena:1999js,Gubser:1998id,Witten:1998ko}, a relationship between gravity and lower-dimensional conformal field theories~\cite{Stephens:1994dx,Susskind:1995hz,Thorn1994}.
The model is exactly solvable in a suitable thermodynamic and conformal limit, where it is conjectured to have a two-dimensional nearly anti-de Sitter space dual, a space-time structure arising close to extremal black holes~\cite{Maldacena:2016hu}.

Majorana fermions by now have several proposed realizations in condensed matter, where they emerge as zero energy bound states in certain superconducting systems~\cite{Kitaev:2000gb,Fu:2008gu,Mourik:2012je,Williams:2012bv,Veldhorst:2012gn,Alicea:2012hz,Leijnse:2012ez,Beenakker:2013jb,Nadj:2013et,Nadj:2014ey,Xu:2015gq,Albrecht:2016cw}. 
Viewing the Majorana fermions in the SYK model as such bound states transforms the system from an intriguing conceptual advance to one with potential physical realizations. 
Existing proposals include Majoranas in a superconducting vortex on the surface of a topological insulator~\cite{Pikulin:2017js}, or multiple Majorana wires connected to a disordered quantum dot~\cite{Chew:2017fn}.
In addition to these condensed-matter proposals, digital quantum simulations on different platforms have been suggested~\cite{Garcia:2017bb}.

As in any condensed-matter setting, such tabletop realizations have a finite number of interacting degrees of freedom (set, e.g., the inserted magnetic flux~\cite{Pikulin:2017js}, or the number of Majorana wires~\cite{Chew:2017fn}).
Working with a finite number $k$ of interacting Majorana modes allows one to consider ``mesoscopic'' features complementary to those in the thermodynamic limit.
These include the level spacing statistics of the many-body eigenenergies, which follow the three Wigner-Dyson symmetry classes---corresponding to random Hermitian matrices with real, complex or quaternionic elements---but in a remarkable eightfold periodic pattern in $k$~\cite{Garcia:2016il,Cotler:2017fx,You:2017jj}.

The origin of this eightfold pattern may be revealed by viewing the SYK model as arising from $k$ Majorana modes living at one end of a one-dimensional time-reversal invariant topological phase~\cite{You:2017jj}.
In the presence of interactions, the topological classification of such systems follows a $\mathbb{Z}_8$ structure, which at the ends of the system is manifested by the emergence of one of the eight (real) Altland-Zirnbauer symmetry classes, depending on the value of $k \mod 8$~\cite{Altland:1997cg,Fidkowski:2011dh}.
Thus, as $k$ is varied, the SYK model is toggled through these classes, which, in turn, imply the eightfold pattern of the three Wigner-Dyson spacing statistics noted in Ref.~\onlinecite{You:2017jj}.

For free fermions, different Altland-Zirnbauer classes show distinct signatures in correlation functions, e.g., in the density of states~\cite{Altland:1997cg}.
These signatures go beyond the three possible Wigner-Dyson level spacing statistics, and characterize the eight real and two complex Altland-Zirnbauer classes. These ten classes together give rise to the so-called ``tenfold way.''

This comparison to free fermion leads to our two main questions:
first, do the eight real Altland-Zirnbauer classes in the SYK model have signatures beyond the simple Wigner-Dyson spacing statistics?
Second, can one go beyond the eight real classes and find SYK realizations of the entire tenfold way? 

In this work, we answer these in the affirmative:
we show that correlation functions carry various signatures of the tenfold way beyond the Wigner-Dyson classification. 
In particular, we show that the eightfold pattern in the purely quartic SYK model is reflected in the strongly interacting analog of the single-particle density of states:
the single-particle spectral function.
Depending on the symmetry class, this displays a peak, a hole, or it is featureless near zero energy.
These are robust characteristics, valid not only for the ground state, but throughout the spectrum and thus for any (including infinite) temperature.
We further show that when bilinear terms are added to the Hamiltonian, the only remaining distinction, namely whether $k$ is even or odd, can be viewed as labeling the two complex Altland-Zirnbauer classes, thus completing the tenfold classification.
These two complex classes, while again identical in terms of their level spacing statistics, are distinguished by the single-particle spectral function, as we show in this work.

In addition to studying spectral features, we also investigate phenomena in the time domain.  
In the thermodynamic limit, the holographic correspondence implies that scrambling, i.e., the spread of locally inserted (into few degrees of freedom) quantum information across the system occurs with maximal efficiency, similar to scrambling in black holes~\cite{Shenker:2014ct,Maldacena:2016gp}.
This maximal scrambling is characterized by an exponential decay of the out-of-time-order correlation function (OTOC) with the maximal Lyapunov exponent~\cite{Kitaev2015,Maldacena:2016hu,Maldacena:2016gp,Maldacena:2016gp_2,Bagrets:2017gk}. 
Working in the mesoscopic regime also provides complementary insights here.
In particular, we find that in contrast to its complete decay in the thermodynamic limit, the $k$-dependent symmetries may dictate that the OTOC  approach a nonzero value at long times, thus linking many-body quantum chaos to the tenfold way.

Our key observation that ties together the various features we find is that, for certain values of $k$, the SYK model supports exact many-body zero modes:  
linear operators local to the SYK model that commute with the Hamiltonian.
Similar objects have  appeared in recent studies of ``strong zero modes'' in one-dimensional interacting Majorana models, spin chains, as well as higher dimensional topologically ordered systems
~\cite{Alicea:2016hp,Fendley:2016ie,Akhmerov:2010dn,Fendley:2012hw,Kemp:2017jy,Goldstein:2012ci,Monthus:2018bu,Else:2017eh,Liu:2017ea}.
That they also arise in the SYK model provides an unexpected link between these emerging directions.

The remainder of this work is organized as follows:
we introduce the SYK model and provide its symmetry classification in Sec.~\ref{sec:model}.
As we show in Sec.~\ref{sec:zero_modes}, the presence of (many-body analogs of) particle-hole and/or chiral symmetries implies the presence of odd-parity many-body zero modes in the SYK model;
depending on the symmetry class, these may have, or be forbidden to have, contributions to Majorana operators, i.e., a single-particle weight.
We turn to the consequences of zero modes and their single-particle weight in Sec.~\ref{sec:correlation_function}, where we consider signatures both in the single-particle spectral function and the OTOC.
In Sec.~\ref{sec:bilinear}, we add symmetry-breaking quadratic terms to the SYK model and show that these lead to the two complex Altland-Zirnbauer classes;
these will also be seen to be distinguished by the presence and absence of zero modes and their energy- and time-domain signatures.
We conclude in Sec.~\ref{sec:conclusion}, where we also give an outlook on experimental perspectives and some generalizations of our results.

\section{Model and symmetry classification}
\label{sec:model}

For the most part, we focus on the SYK Hamiltonian with random four-body interactions~\cite{Sachdev:1993hv,Kitaev2015,Maldacena:2016hu}
\begin{equation}
 H = \sum_{t=0}^{k-1} \sum_{s=0}^{t-1} \sum_{r=0}^{s-1} \sum_{q=0}^{r-1}  J_{qrst} \gamma_q \gamma_r \gamma_s \gamma_t
 \label{eq:hamiltonian_SYK}
\end{equation}
between $k$ Majorana modes that obey the anticommutation relation $ \left\lbrace \gamma_q ,\gamma_r \right\rbrace = 2 \delta_{qr}$ with $\gamma_q^\dagger = \gamma_q$.
(A Hamiltonian with additional terms is considered in Sec.~\ref{sec:bilinear}.)
The random interaction $J_{qrst}$, with zero mean $\langle J_{I} \rangle = 0$, defines the system's only energy scale $J$ via its variance
\begin{align}
 \langle J_{I} J_{I'} \rangle = \frac{3!}{k^3} J^2\,\delta_{I,I'} ,
 \label{eq:statistics_J}
\end{align}
where $\langle \ldots \rangle$ denotes averaging over different realizations of the random interaction $J_{qrst}$, with combined indices $I=(qrst)$.

\begin{table}
 \begin{tabular}{c|c|c|c}
 \toprule
  $k \mod 8$ & ~$T_+^2$~ & ~$T_-^2$~ & ~Cartan label \\
  \colrule
  0	& $+1$ & $0$  & AI \\
  1	& $+1$ & $+1$ & BDI \\
  2	& $0$  & $+1$ & D \\
  3	& $-1$ & $+1$ & DIII \\
  4	& $-1$ & $0$  & AII \\
  5	& $-1$ & $-1$ & CII \\
  6	& $0$  & $-1$ & C \\
  7	& $+1$ & $-1$ & CI\\
  \botrule
 \end{tabular}
 \caption{Eightfold symmetry classification of the SYK model.
 Time-reversal ($T_+$) and particle-hole symmetry ($T_-$) may be absent (labeled by $0$) or present, labeled by the squares $T_+^2= \pm 1$ and $T_-^2 = \pm 1$.
 We give the corresponding Cartan labels in the last column.
 The correspondence between the sign structure and the Cartan labels follows the same pattern as for free fermions.
 However, while for free fermions $T_\pm$ and their sign structure represent physically different symmetries, in the SYK model they arise from the same physical symmetry $T \gamma_q T^{-1} = \gamma_q$ and its properties, especially its interplay with parity, which changes with the number $k$ of Majorana fermions.
 }
 \label{tab:cartan}
\end{table}

Depending on the number of Majorana modes contributing to the Hamiltonian, the SYK model realizes one of the eight real symmetry classes of the Altland-Zirnbauer classification, a result we obtain below following Ref.~\onlinecite{Fidkowski:2011dh} and summarize in Table~\ref{tab:cartan}.
(This realization of the Altland-Zirnbauer classes is distinct from that arising in the context of supersymmetric generalizations of the SYK model~\cite{Fu:2017hl,Kanazawa:2017be,Li:2017bi,Garcia:2018jz}.)
The Altland-Zirnbauer classification originates in random-matrix theory and, in its original context, determines universal spectral properties of single-particle Hamiltonians~\cite{Altland:1997cg}.
For a single-particle Hamiltonian $\mathcal{H}$ (or more generally its irreducible subblock with respect to unitary symmetries) two independent antiunitary operators $T_\pm$ may exist that commute (time-reversal symmetry $[ T_+, \mathcal{H}] = 0$) or anticommute (particle-hole symmetry $\lbrace T_- ,\mathcal{H} \rbrace = 0$) with $\mathcal{H}$.
When the operators $T_\pm$ are present, they either square to $T_\pm^2 = +1$ or $T_\pm^2 = -1$.
This gives nine distinct possibilities for the antiunitary symmetries $\mathcal{H}$ may possess: absence of $T_\pm$, presence with $T_\pm^2 = +1$ or presence with $T_\pm^2 = -1$.
When both symmetries are absent, their combination, the unitary chiral symmetry $Z=T_+ T_-$ with $\lbrace Z, \mathcal{H} \rbrace = 0$, can still be present.
This results in a total of ten different symmetry classes.
Eight of these ten classes have a reality condition: the Hamiltonian and its complex conjugate are related via one or both of the antiunitary operators $T_+$ and $T_-$;
thus, we refer to these classes as real symmetry classes. Conversely, the two remaining classes we refer to as complex symmetry classes.

Many-body systems may also be classified according to their antiunitary symmetries~\cite{Fidkowski:2010ko,Fidkowski:2011dh,Pollmann:2010ih,Turner:2011gp,Chen:2011fp}.
For example, \citeauthor{Fidkowski:2011dh} considered gapped one-dimensional fermion systems with $k$ Majorana fermion end modes $\gamma_q$ satisfying $T \gamma_q T^{-1} = \gamma_q$ for some antiunitary $T$. 
They showed that upon restricting considerations to the Hilbert space for only these Majoranas, the properties of $T$ are set only by $k$.
Depending on $k\mod 8$, $T$ may preserve fermion parity, in which case we denote it by $T=T_+$, or flip it, which we denote by $T=T_-$.
(For odd $k$, an additional Majorana at infinity that is necessary for a valid fermion Hilbert space allows one to construct both $T_+$ and $T_-$, see below.) The square of $T_\pm$, when present, also depends on $k$. 
There is thus a $k$-dependent symmetry classification of Hamiltonians for $k$ Majorana end modes. 
As \citeauthor{Fidkowski:2011dh}  also showed, this emergent eightfold structure provides an incarnation of the eightfold (real) Altland-Zirnbauer classification.

While in \citeauthor{Fidkowski:2011dh}'s work the symmetry $T \gamma_q T^{-1} = \gamma_q$ arose at the end of a one-dimensional time-reversal-invariant system (with Majorana fermions at the other end satisfying $T \gamma_q T^{-1}= -\gamma_q$), such a symmetry is also a symmetry of the SYK Hamiltonian~\eqref{eq:hamiltonian_SYK}: $THT^{-1}=H$.
In fact, any antiunitary symmetry $T$ of the SYK model can be shown to imply $T \gamma_q T^{-1}= \pm \gamma_q$ (with $q$-independent signs):
phases other than $\pm 1$ are incompatible with unitary operators $\gamma_q^\dagger = \gamma_q$ and mixing signs or indeed different $\gamma_q$ does not result in a symmetry for all realizations of the random couplings $J_{qrst}$.
We can therefore directly apply this classification scheme to the SYK model~\cite{You:2017jj}.
Here we provide a summary of the main results formulated in terms of the $T_\pm$ notation introduced above.
To make the work self contained, we show in Appendix~\ref{sec:symmetries} how higher-dimensional Clifford algebras~\cite{de1986field,Brauer:1935ji,Pais:1962is,Kennedy:1981eu} or an explicit construction of $T_\pm$ as a product of Majorana operators may be used to obtain these results, with the former providing an alternative approach to that in Ref.~\onlinecite{Fidkowski:2011dh}.
To connect with the symmetries associated to Altland-Zirnbauer classes in free fermion systems, we refer to $T_+$ as time-reversal and $T_-$ as particle-hole symmetry;
this allocation can be motivated, e.g., by observing that similarly to time-reversal in the free fermion case, it is the $T_+$ properties that set the level spacing statistics~\cite{Garcia:2016il,You:2017jj,Cotler:2017fx}. (In contrast, $T_-$ will be seen to set features beyond Wigner-Dyson, as we shall explain.)

We first discuss the case when $k$ is even.
In this case the fermion parity is simply the product of all the Majorana fermions~\cite{Kitaev:2000gb},
\begin{equation}
 P = i^{k/2} \gamma_0 \gamma_1 \cdots \gamma_{k-1}
 \label{eq:parity_even_k}
\end{equation}
where the phase is chosen such that $P^\dagger = P$.
This expression can be viewed as the product of single fermion parities $i \gamma_j \gamma_k = (-1)^{d_{jk}^\dagger d_{jk}}$ where $d_{jk} = (\gamma_j + i \gamma_k)/2$ is an ordinary (``complex'') fermion constructed from $\gamma_j$ and $\gamma_k$.
When $k=4n$, $P$ consists of Majorana operators multiplied by a real number and therefore $T\gamma_q T^{-1}= \gamma_q$ combined with antiunitary of $T$ implies $T P T^{-1}=P$; thus, $T=T_+$ in this case.
Conversely, when $k=4n+2$, $P$ consists of Majorana operators multiplied by a purely imaginary number, and thus $T=T_-$ in this case.
We obtain the square $T_\pm^2$ by writing $T$ as a product of complex conjugation and $k/2$ Majorana operators~\cite{Fidkowski:2010ko}, which are a higher-dimensional generalization of charge conjugation matrices~\cite{Pais:1962is} (see Appendix~\ref{sec:symmetries} for details).

For an odd number of Majorana modes, which can for example be realized at topological defects (e.g,. a vortex) or at the boundary of a higher-dimensional system, a fermion parity operator cannot be directly defined as it requires an even number of Majorana modes.
Instead, we need to introduce an additional decoupled Majorana fermion $\gamma_\infty$, which does not enter the Hamiltonian and may be thought of as residing at a different boundary or defect.
The parity operator then takes the form 
\begin{equation}
 P = i^{(k+1)/2} \gamma_0 \gamma_1 \cdots \gamma_{k-1}\gamma_\infty .
 \label{eq:parity_odd_k}
\end{equation}
It turns out useful to define the Hermitian and unitary operator $Z$
\begin{equation}
 Z = i^{(k-1)/2} \gamma_0 \gamma_1 \cdots \gamma_{k-1} .
 \label{eq:definition_z}
\end{equation}
This object may be informally thought of as a composite Majorana fermion complementary to $\gamma_\infty$, since fermion parity is now given by $P = i Z \gamma_\infty$, and
$Z$ anticommutes with $P$ and $\gamma_\infty$. $Z$, however, commutes with all $\gamma_{q\neq\infty}$, and consequently with the Hamiltonian.
As the notation suggests, this operator has an interpretation as an operator for chiral symmetry, as we show below.
Note that the states $\ket{\psi^p}$ and $\ket{ Z \psi^p}$ have the same energy eigenvalue (since $[ Z, H ] = 0$) but different parity eigenvalue (since $\lbrace Z, P \rbrace = 0$), which implies that the two parity sectors are always degenerate for odd $k$.

For odd $k$, it is always possible to find both operators $T_\pm$ that act as time-reversal and particle-hole symmetry on $\gamma_{q\neq \infty}$.
Once we find an operator $T_+$ that commutes with all $\gamma_{q\neq\infty}$ and with $P$, we can construct the particle-hole symmetry $T_- = T_+^{-1} Z$ with $\lbrace T_- , P \rbrace = T_+^{-1} \lbrace Z , P \rbrace = 0$.
Thus viewing $Z=T_+ T_-$ motivates interpreting $Z$ as a chiral symmetry.
Analogously to the case with even $k$, the squares $T_\pm^2=\pm 1$ may be obtained by constructing the operators in terms of Majorana operators or by using the Clifford algebra structure; see Appendix~\ref{sec:symmetries}.
We summarize the results in Table~\ref{tab:cartan}, together with the corresponding Cartan labels~\cite{Altland:1997cg}.

As we mentioned above, based on $T_+$ alone, the system may be thought to belong to one of three Wigner-Dyson classes:
$T_+$ absent (unitary),  $T_+^2=+1$ (orthogonal), or $T_+^2=-1$ (symplectic).
This has important consequences for the level spacing statistics~\cite{Wigner:1951bx,Wigner:1958gl,Dyson:1962ib} of the SYK Hamiltonian~\eqref{eq:hamiltonian_SYK}, which have been shown to follow the corresponding Gaussian ensembles~\cite{Garcia:2016il,You:2017jj,Cotler:2017fx}.

It is known already for single-particle problems that the different level spacing statistics are not sufficient to distinguish the Altland-Zirnbauer classes~\cite{Altland:1997cg}.
Instead, one needs to consider other observables, such as the single-particle spectral function, which simplifies to the average level density  in the noninteracting case.
Key distinguishing features include the possibility, or the impossibility, of robust zero energy modes for classes with particle-hole and/or chiral symmetries~\cite{Ivanov:2002ho}.
A particular example pertains to classes D and C:
while in terms of their level spacing statistics the two may look identical, it is only the single-particle Hamiltonian of the former that can support a robust nondegenerate zero mode protected by particle-hole symmetry~\cite{MehtaBook}.
For the latter, the fact that particle-hole symmetry squares to minus unity implies level degeneracy at, and only at, zero energy, precluding the presence of a robust zero mode.
The spectral density, instead, displays a spectral hole near zero energy~\cite{Altland:1997cg}.
As these features follow from the single-particle Hamiltonian, the resulting zeros modes may be viewed as single-particle zero modes.

Based on the existence of such a taxonomy of features tied to the symmetry classification of noninteracting fermions, it is tempting to ask whether an analogous taxonomy of features, such as symmetry-guaranteed or forbidden many-body zero modes, may emerge for the Altland-Zirnbauer classes of the---interaction only---SYK model.
This is what we turn to in the next section.

\section{Zero modes and matrix overlaps}
\label{sec:zero_modes}

Zero modes are local (linear) operators that commute with the Hamiltonian~\cite{Fendley:2016ie,Goldstein:2012ci,Monthus:2018bu}. 
(For a zero-dimensional system such as the SYK model, by local we mean local to the Hamiltonian, i.e., not involving external degrees of freedom such as $\gamma_\infty$.)
They have been discussed in other contexts, such as anyonic excitations~\cite{Alicea:2016hp,Fendley:2016ie,Akhmerov:2010dn}, spin chains~\cite{Fendley:2012hw,Kemp:2017jy}, and certain interacting Majorana models~\cite{Goldstein:2012ci,Monthus:2018bu,Else:2017eh,Liu:2017ea}.
But, to the best of our knowledge, the idea that the SYK model could support zero modes, and the connection of these to the above symmetry classification, has not yet been considered.
In Ref.~\onlinecite{Fendley:2016ie}, only operators that do not commute with a discrete symmetry of the Hamiltonian were considered zero modes; here, following Ref.~\onlinecite{Goldstein:2012ci}, we use a broader definition that distinguishes between even-parity zero modes that commute with the parity operator $P$ and odd-parity zero modes that anticommute with $P$.

Zero modes are eigenoperators $\mathcal{O}_i$ of the Hamiltonian
\begin{equation}
 \left[ H ,\mathcal{O}_i \right] = \lambda_i \mathcal{O}_i
 \label{eq:eigenoperators}
\end{equation}
with $\lambda_i = 0$.
Such zero modes are many-body generalizations of zero-energy eigenstates of single-particle Hamiltonians.
Since the system's total energy does not depend on the occupancy of these eigenstates, the operators that create or annihilate them must commute with the second-quantized Hamiltonian.
As the SYK model is an interaction-only model without an underlying single-particle picture, we use the term ``zero modes'' to refer only to the many-body operators introduced above.

Previous approaches for Majorana systems solved Eq.~\eqref{eq:eigenoperators} by expanding $\mathcal{O}_i$ into an operator basis, 
obtaining 
matrices that are exponentially large (in the number of Majoranas) to be
diagonalized to obtain solutions with $\lambda_i = 0$~\cite{Goldstein:2012ci}.
Although all zero modes can be obtained within this approach for each realization of the Hamiltonian, further information, e.g., the degeneracy of the zero modes (a key factor, e.g., if one wishes to consider zero mode based qubits), cannot be argued for on general grounds away from the weakly interacting limit.

Instead of expanding Eq.~\eqref{eq:eigenoperators} in an operator basis, below we introduce a direct method that not only delivers the zero modes, and indeed all eigenmodes, in terms of the eigenstates of the system, but immediately accounts for their degeneracies for arbitrary values and forms of interactions. 

Our key observation is that if $H$ is diagonalized as
\begin{equation}
 H= \sum_\mu \varepsilon_\mu \ket{\psi_\mu} \bra{\psi_\mu},
\end{equation}
then the eigenoperator equation is solved by
\begin{align}
 \mathcal{O}_i \equiv \mathcal{O}_{\mu\nu} = \ket{\psi_\mu} \bra{\psi_\nu} , & &
 \lambda_i \equiv \lambda_{\mu\nu} = \varepsilon_\mu - \varepsilon_\nu.
 \label{eq:eigenoperators_general}
\end{align}
All the solutions arise this way, as seen by noting that for Hilbert space dimension $M$, we have $M^2$ linearly independent operators $\mathcal{O}_{\mu\nu}$, which exhaust the dimension of the operator Hilbert space [and hence the possibilities for $\mathcal{O}_i$ in Eq.~\eqref{eq:eigenoperators}].
While the above observation is general, more care needs to be taken to ensure that the zero modes are local. For the SYK model, possible nonlocality arises only for odd $k$, due to the presence of the external degree of freedom $\gamma_\infty$ in the operator algebra.  

\subsection{Even $k$}

For even $k$, there are $2^{k-1}$ even-parity eigenoperators that preserve parity
\begin{align}
 \mathcal{O}_{p\mu\nu}^e = \ket{\psi_{\mu}^p} \bra{\psi_\nu^p}, & &
 \left[ H ,\mathcal{O}_{p\mu\nu}^e \right] = (\varepsilon_\mu^p - \varepsilon_\nu^p ) \mathcal{O}_{p\mu\nu}^e
 \label{eq:zero_modes_ee}
\end{align}
labeled by the parity $p=\pm1$ and $H \ket{\psi^p_\mu} =  \varepsilon_\mu^p \ket{\psi_\mu^p}$ with $\mu,\nu \in [1,\ldots, 2^{k/2-1} ]$.
The operators $\mathcal{O}_{p\mu\mu}^e$ are trivially zero modes of the Hamiltonian, as they are simply projectors on the eigenstate $\ket{\psi_\mu^p}$.
In class AII, another even zero mode is present due to Kramers degeneracy with $T_+^2 = -1$:
for each state $\ket{\psi_\mu^p}$, the orthogonal state $\ket{\psi_{\mu'}^p} = T_+ \ket{\psi_\mu^p}$ has the same energy, resulting in $\mathcal{O}_{p\mu\mu'}^e$ being a zero mode.

In addition to the even eigenoperators, there are $2^{k-1}$ odd-parity eigenoperators
\begin{align}
 \mathcal{O}_{p\mu\nu}^o = \ket{\psi_{\mu}^p} \bra{\psi_\nu^{-p}} , & &
 \left[ H ,\mathcal{O}_{p\mu\nu}^o \right] = (\varepsilon_\mu^p - \varepsilon_\nu^{-p} ) \mathcal{O}_{p\mu\nu}^o
 \label{eq:zero_modes_eo}
\end{align}
that change the parity of a state.
In classes AI and AII, generally $\varepsilon_\mu^p \neq \varepsilon_\nu^{-p}$ for all $\mu,\nu$, so no odd zero modes are present. %
In classes C and D, however, the two parity sectors are degenerate $\varepsilon_\mu^p = \varepsilon_\mu^{-p}$ due to the presence of a particle-hole symmetry $T_-$:
the states $\ket{\psi_\mu^+}$ and $\ket{\psi_\mu^{-}} \equiv T_- \ket{\psi_\mu^+}$ are orthogonal and have the same energy.
Thus, all $\mathcal{O}_{p\mu\mu}^o$ are odd-parity zero modes of the Hamiltonian (and these are the only odd-parity zero modes due to the absence of other degeneracies between the parity sectors).
This is a nontrivial result:
while the presence of even-parity zero modes is obvious from this construction, the presence of odd-parity zero modes is not generally expected to happen for an even number of Majorana fermions.
It is a key finding of this work that the symmetry classification of the SYK model ensures the presence of odd zero modes in classes C and D.

The contribution of the zero modes to correlation functions depends on the symmetry class.
To see this, we expand the Majorana operators in terms of the odd eigenoperators $\mathcal{O}^o$ and split the sum
\begin{equation}
 \gamma_q = \underbrace{
 \sum_{p \mu} u_{q,p\mu\mu} \mathcal{O}^o_{p\mu\mu}}_{\equiv A_q}
  +  \underbrace{
 \sum_{ \substack{p\mu\nu \\ \mu \neq \nu}} u_{q,p\mu\nu} \mathcal{O}^o_{p\mu\nu} }_{\equiv B_q}
\label{eq:gamma_q_even}
\end{equation}
into a zero mode term $A_q$ with $[ A_q , H ] = 0$ and the rest, $B_q$,  with $[B_q , H ] \neq 0$.
The zero mode coefficients are  
\begin{equation}\label{eq:keven_Aqcoeff}
 u_{q,p\mu\mu} = \bra{\psi^p_\mu} \gamma_q \ket{\psi^{-p}_\mu} = \bra{\psi^p_\mu} \gamma_q T_- \ket{\psi^{p}_\mu},
\end{equation}
where the last equality holds up to a sign that can be neglected in the further analysis.
The presence of the antiunitary operator $\gamma_q T_-$ places restrictions on when $u_{q,p\mu\mu} \neq 0$ is possible.
In particular, when $(\gamma_q T_-)^2 =-1$, this guarantees $u_{q,p\mu\mu}=0$ by the same mechanism as that behind Kramers' theorem. 

Using $[ T_-,\gamma_q ] = 0$, we get $(\gamma_q T_-)^2 = T_-^2$ and therefore
\begin{equation}
 u_{q,p\mu\mu} =
 \begin{cases} 0 & \text{class C} \\ \text{nonzero} & \text{class D.} \end{cases}
\end{equation}
Thus, we generally have a zero-mode contribution to the Majorana operators in class D, but not in class C.
The structure we have uncovered shows an emergent Altland-Zirnbauer pattern of   features that can be viewed as many-body incarnations of those in noninteracting systems:  %
In particular, classes C and D, which are identical as far as energy spacing statistics go, are clearly distinguished by the presence or absence of the zero mode contribution $A_q$.
The absence of odd zero modes for classes AI and AII is again analogous to the absence of zero modes in these classes in the noninteracting case.

\subsection{Odd $k$}
\label{sec:zero_modes_odd}

For odd $k$, the two parity sectors are related via \mbox{$\ket{\psi_\mu^p} = Z \ket{\psi_{\mu}^{-p}}$.}
Since $Z$ commutes with the Hamiltonian, the two parity sectors are guaranteed to be degenerate, $\varepsilon_\mu^p = \varepsilon_\mu^{-p} \equiv \varepsilon_\mu$.
This implies that the previously defined operators $\mathcal{O}^{e/o}_{p\mu\nu}$ are not local to the SYK model, 
i.e., they include the operator $\gamma_\infty$: 
If $\mathcal{O}^e_{p\mu\nu} = \ket{\psi_\mu^p} \bra{\psi_\nu^p}$ were local, it would commute with $Z$ (since $[\gamma_{q\neq\infty}, Z]=0$);
this would imply $\ket{\psi_\mu^p} \bra{\psi_\nu^p}=Z\ket{\psi_\mu^p} \bra{\psi_\nu^p}Z = \ket{\psi_\mu^{-p}} \bra{\psi_\mu^{-p}}$, which is not possible since these operators project on opposite parity subspaces.
Instead, the $2^{k-1}$ different linear combinations of even operators
\begin{subequations}
 \label{eq:zero_modes_oe}
\begin{align}
 \mathcal{O}_{\mu\nu}^e &= \ket{\psi_{\mu}^+} \bra{\psi_\nu^+} + \ket{\psi_{\mu}^-} \bra{\psi_\nu^-} \\
 \left[ H ,\mathcal{O}_{\mu\nu}^e \right] &= (\varepsilon_\mu - \varepsilon_\nu ) \mathcal{O}_{\mu\nu}^e
 \end{align}\end{subequations}
and $2^{k-1}$ odd eigenoperators
\begin{subequations}
 \label{eq:zero_modes_oo}
\begin{align}
 \mathcal{O}_{\mu\nu}^o &= \ket{\psi_{\mu}^+} \bra{\psi_\nu^{-}} + \ket{\psi_{\mu}^-} \bra{\psi_\nu^{+}} \\
 \left[ H ,\mathcal{O}_{\mu\nu}^o \right] &= ( \varepsilon_\mu - \varepsilon_\nu ) \mathcal{O}_{\mu\nu}^o
\end{align}\end{subequations}
that commute with $Z$, and thus do not include $\gamma_\infty$, span the local Hilbert space.
Due to parity degeneracy, both $\mathcal{O}_{\mu\mu}^e$ and $\mathcal{O}_{\mu\mu}^o$ are zero modes.
When $T_+^2=-1$, as in classes DIII and CII, Kramers degeneracy results in more even and odd zero modes, $\mathcal{O}_{\mu\mu'}^e$ and $\mathcal{O}_{\mu\mu'}^o$, involving time-reversed partners $\ket{\psi^p_{\mu'}} \neq \ket{\psi^p_\mu}$ with $\varepsilon_\mu = \varepsilon_{\mu'}$, defined via $\ket{\psi^+_{\mu'}} \equiv T_+ \ket{\psi^+_\mu}$.

As in the $k$ even case, the Majorana operators can be split up a zero mode part and the rest,
\begin{equation}
 \gamma_q = \underbrace{\sum_{\substack{\mu\nu \\ \varepsilon_\mu = \varepsilon_\nu} } u_{q,\mu\nu} \mathcal{O}^o_{\mu\nu}}_{\equiv A_q} +  \underbrace{\sum_{ \substack{\mu\nu \\ \varepsilon_\mu \neq \varepsilon_\nu}} u_{q,\mu\nu} \mathcal{O}^o_{\mu\nu} }_{\equiv B_q}
\label{eq:gamma_q_odd}
\end{equation}
with
\begin{align}\label{eq:eq:kodd_Aqcoeff}
 u_{q,\mu\nu}
 &= \frac{1}{2} \sum_p \bra{\psi_\mu^p} \gamma_q \ket{\psi_\nu^{-p}}
  = \frac{1}{2} \sum_p \bra{\psi_\mu^p} \gamma_q Z \ket{\psi_\nu^p},
\end{align}
where the factor of $1/2$ is due to $\mathrm{tr} [ \mathcal{O}^{e}_{\mu\nu} ] = 2\delta_{\mu\nu}$.
Only those elements $u_{q,\mu\nu}$ with $\varepsilon_\mu = \varepsilon_\nu$ contribute to $A_q$.
Two different symmetries demand opposite-parity states of the same energy:
chiral symmetry with matrix elements
\begin{equation}
 u_{q,\mu\mu} = \frac{1}{2} \sum_p \bra{\psi_\mu^p} \gamma_q Z \ket{\psi_\mu^p} ,
\end{equation}
and particle-hole symmetry with
\begin{align}
 u_{q,\mu \mu'}
  &= \frac{1}{2} \sum_p \bra{\psi_\mu^p} \gamma_q \ket{\psi_{\mu'}^{-p}}
  =  T_-^2 \bra{\psi_\mu^+} \gamma_q T_- \ket{\psi_{\mu}^{+}} .
\end{align}
The zero-mode contribution to $\gamma_q$ can be rewritten using the notation introduced above,
\begin{equation}
 A_q = \sum_{\mu} u_{q,\mu\mu} \mathcal{O}^o_{\mu\mu} + \sum_{\mu} u_{q,\mu \mu'} \mathcal{O}^o_{\mu \mu'}.
\end{equation}
The second contribution, involving matrix elements $u_{q,\mu\mu'}$, is zero when $(\gamma_q T_-)^2 = T_-^2 = -1$, i.e.,
\begin{equation}
 u_{q,\mu\mu'} =	
 \begin{cases} 0 & \text{classes CI, CII} \\ \text{nonzero} & \text{classes BDI, DIII.} \end{cases}
\end{equation}
Further, when $T_+^2 = + 1$, i.e., in classes CI and BDI, $T_+\ket{\psi_\mu^p}$ and $\ket{\psi_\mu^p}$ only differ by a phase, such that each parity sector is nondegenerate.
Accordingly, the states $Z \ket{\psi_\mu^p}$ and $T_-\ket{\psi_\mu^p}$ also differ only by a phase, thus $| \bra{\psi_{\mu}^p} \gamma_q Z \ket{\psi_\mu^p} |
 = | \bra{\psi_{\mu}^p} \gamma_q T_- \ket{\psi_\mu^p} |$ and therefore
\begin{equation}
u_{q,\mu\mu}=\begin{cases} 0 & \text{class CI} \\ \text{nonzero} & \text{class BDI.} \end{cases}
\end{equation}
Thus, in class CI, all contributions to $A_q$ are zero.
When $T_+^2 = -1$, i.e., in classes CII and DIII, $Z \ket{\psi_\mu^p}$ and $T_- \ket{\psi_\mu^p}$ are orthogonal, thus $\bra{\psi_{\mu}^p} \gamma_q Z \ket{\psi_\mu^p}$ is independent of $\bra{\psi_{\mu}^p} \gamma_q T_- \ket{\psi_\mu^p}$ and hence $u_{q,\mu\mu}\neq 0$ in general.
Together with the results for even $k$, we summarize the symmetry constraints on the matrix elements in Table~\ref{tab:overlap}.

\begin{table}
 \begin{tabular}{c|c|c|c}
 \toprule
  $k \mod 8$	& $\bra{\psi^p_\mu} \gamma_q T_- \ket{\psi_\mu^p}$ & $\bra{\psi_\mu^p} \gamma_q Z \ket{\psi_\mu^p}$ & Cartan label \\
  \colrule
  0	& --			& --			& AI	\\
  1	& nonzero	& nonzero	& BDI	\\
  2	& nonzero	& --			& D	\\
  3	& nonzero	& nonzero	& DIII	\\
  4	& --			& --			& AII	\\
  5	& $0$		& nonzero	& CII	\\
  6	& $0$		& --			& C	\\
  7	& $0$		& $0$		& CI	\\
  \botrule
 \end{tabular}
 \caption{Expectation value of the parity-conserving operators $\gamma_q T_-$ and $\gamma_q Z$ for eigenstates with a well-defined parity $p$.
A nonzero value of at least one of these matrix elements is required for a nonvanishing zero mode contribution $A_q$ to the Majorana operator $\gamma_q$. 
The cases where $Z$ or $T_-$ do not exist are marked by --. 
 When the $Z$ and/or $T_-$ exist, the overlap might be zero due to symmetry restrictions (marked by $0$); otherwise, it is generally nonzero.
 }
 \label{tab:overlap}
\end{table}

\section{Correlation functions}
\label{sec:correlation_function}

We now turn to discussing the consequences of the Altland-Zirnbauer classes and the SYK zero modes on various correlation functions. 
As the zero modes relate to pairs of equal-energy eigenstates, their main effect is expected to be at small energies (or long times in the time domain), though with no restriction for this energy to be relative to the ground state, or even a thermal state. 
In the energy domain, we focus on the behavior of Majorana single-particle spectral functions, while in the time domain, in addition to commenting on the consequences of our findings on the spectral function, we investigate how the zero modes influence the behavior of the OTOC.
The OTOC is related to the effect of one observable on another observable at a later time, and shows the butterfly effect in many-body chaotic systems~\cite{Hayden:2007bc,Sekino:2008im,Patel:2017fq,Rozenbaum:2017br,Nosaka:2018iy,Das:2018km}. 

\subsection{Single-particle correlations}

The Majorana single-particle spectral function is a quantity accessible in solid-state realizations of the SYK model through tunneling experiments~\cite{Pikulin:2017js,Chew:2017fn}.
In terms of the retarded single-particle Green's function $C^+_{qq} (\omega)$, it reads
\begin{equation}
 A (\omega) = - \frac{1}{k}\,\frac{1}{\pi} \mathrm{Im} \sum_q C^+_{qq} (\omega) ,
 \label{eq:spectral_function}
\end{equation}
where the single-particle Green's function can be expressed in K\"{a}ll\'{e}n-Lehmann representation as~\cite{AltlandSimons}
\begin{align}
 C^+_{qr} (\omega) =& \frac{1}{\mathcal{Z}} \sum_{\mu\nu p} \frac{\bra{\psi_\mu^p} \gamma_q \ket{\psi_{\nu}^{-p}} \bra{\psi_{\nu}^{-p}} \gamma_r \ket{\psi_\mu^p}}{ \omega + \varepsilon_{\mu}^p - \varepsilon_{\nu}^{-p} + i \eta}  \label{eq:correlation_function} \\
 & \times \left( e^{-\beta \varepsilon_\mu^p} + e^{-\beta \varepsilon_{\nu}^{-p}} \right), \nonumber
\end{align}
with the inverse temperature $\beta$ and infinitesimal $\eta>0$ that corresponds to level broadening. 
As we show below, as a consequence of the zero modes, the spectral function has either a peak, a hole or is featureless near $\omega = 0$.

\subsubsection{Spectral function at zero temperature}

At zero temperature, the spectral function~\eqref{eq:spectral_function} simplifies to
\begin{align}
 A (\omega) =& \frac{1}{k} \sum_q \sum_{\mu p}  \left[ \delta (\omega + \varepsilon_{0}^p - \varepsilon_{\mu}^{-p} ) + \delta (\omega + \varepsilon_{\mu}^{-p}-\varepsilon_{0}^p ) \right] \nonumber \\
 & \times \left| \bra{\psi_\mu^p} \gamma_q \ket{\psi_0^{-p}} \right|^2,
 \label{eq:spectral_function_T0}
\end{align}
i.e., only the overlap with the ground state enters [$\nu\to 0$ in Eq.~\eqref{eq:correlation_function}].
To resolve the spectral function at small energies, we focus on $\omega \sim \Delta_0$, the average first excitation energy $\Delta_0 = \langle \varepsilon_1^p-\varepsilon_0^p \rangle$ (this is independent of $p$).
For large $k$, one may express $\Delta_0$ analytically using approximations~\cite{Bagrets:2016gm,Garcia:2016il} for the single-particle density of states at $\omega \sim \varepsilon_0$; however, since we are especially interested in a moderate number of Majoranas, we employ a more pragmatic numerical approach and obtain $\Delta_0$ by numerically averaging over a large ensemble of the random interactions $J_{qrst}$.
When computing the correlation function~\eqref{eq:spectral_function_T0}, we replace the delta function by a Lorentz function $\delta (x) = \lim_{\eta \to 0} \eta/[\pi(\eta^2 + x^2)]$, with $\eta$ chosen as $\eta=c\Delta_0$ with constant $c$ much smaller than unity.

\begin{figure}
 \includegraphics[width=\linewidth]{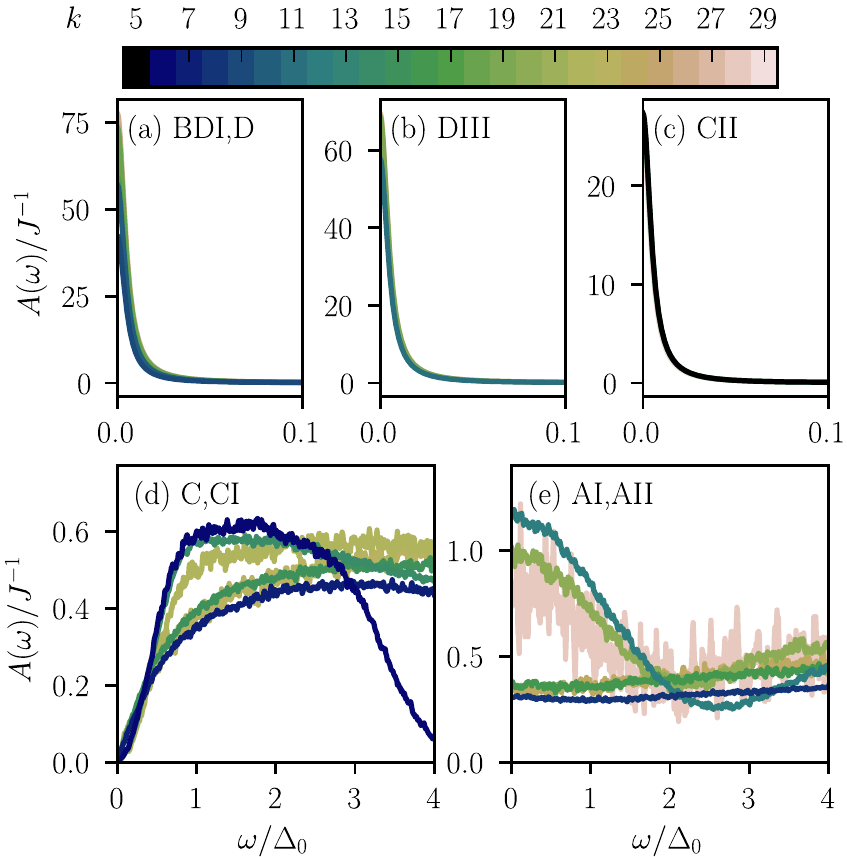}
 \caption{Spectral function at zero temperature vs $\omega/\Delta_0$.
 The level broadening is $\eta = 0.005 \Delta_0$.
 Panels~(a)--(c) show symmetry classes where the spectral function exhibits a peak at zero energy (with different peak sizes), panel~(d) shows classes with a hole, and~(e) shows featureless behavior.
 Different colors denote different $k$.
 }
 \label{fig:dos_at_zero}
\end{figure}

In Fig.~\ref{fig:dos_at_zero}, we show the numerically evaluated spectral function at zero temperature, averaged over a large ensemble [ranging from $2^6$ to $2^{12}$ realizations of $J_{qrst}$ in panels (a)--(c), and from $2^{11}$ to $2^{18}$ realizations in panels (d) and (e)].
Panels~(a)--(c) show cases with $A_q \neq 0$.
The corresponding nonvanishing $\bra{\psi_0^p} \gamma_q \ket{\psi_0^{-p}}$ overlaps [cf. Eqs.~\eqref{eq:keven_Aqcoeff} and \eqref{eq:eq:kodd_Aqcoeff}] translate into a nonzero weight for $\delta(\omega)$: the spectral function displays a $\omega=0$ peak. 

Classes C and CI host zero modes, but their contribution to $\gamma_q$ is zero, $A_q = 0$, as shown in the previous section.
In both classes, the spectrum is the same for both parities, thus level repulsion is present across the two parity sectors.
Together with $A_q=0$, this results in the spectral function having a hole at $\omega = 0$, panel~(d). This is again similar to the noninteracting case, specifically to the spectral hole for classes C and CI at zero energy.
Up to energies $\omega \sim \Delta_0$, the spectral function follows the gap statistics, $\varepsilon_1 - \varepsilon_0$, i.e., the Gaussian unitary ensemble (C) and Gaussian orthogonal ensemble (CI), once we unfold the energies $\omega \to \xi (\omega)$ to have the same mean level spacing everywhere in the spectrum, i.e., a constant density of states (cf.\ Refs.~\onlinecite{You:2017jj,Guhr:1998bg} for the unfolding procedure).
In particular, for $\xi \ll \Delta_0$, we have a power law scaling $A(\xi) \propto \xi^\beta$ with $\beta=1$ for class CI and $\beta=2$ for class C, in complete analogy with the spectral density of the corresponding symmetry classes in the noninteracting case~\cite{Altland:1997cg}.

These features distinguish the (real) classes BDI, D, DIII, CII, C and CI from the two real Wigner Dyson classes AI (orthogonal) and AII (symplectic): as shown in Fig.~\ref{fig:dos_at_zero}~(e), the latter two classes display no peak nor hole for $\omega$ near zero. 
This featureless $\omega \ll \Delta_0$ behavior 
results from the two parity sectors being independent:
although the energies in each parity sector follow the level statistics from the Gaussian orthogonal ensemble (AI) or Gaussian symplectic ensemble (AII) individually, the differences $\varepsilon_\mu^p - \varepsilon_\nu^{-p}$ do not---there is no level repulsion between the parity sectors, such that we do not observe the typical power laws at low energies.

\subsubsection{Spectral function at infinite temperature}

The equal-energy matrix elements associated with the nonzero single-particle contribution $A_q\neq 0$ are present for generic eigenstates, not only the groundstate. 
In particular, the spectral function displays the peaks and holes following the same pattern relative to any eigenstate.
The pattern is therefore also reflected in the infinite-temperature spectral function
\begin{align}
 A (\omega) =& \frac{1}{k} \frac{2}{M} \sum_q \sum_{\mu\nu p} \left| \bra{\psi_\mu^p} \gamma_q \ket{\psi_\nu^{-p}} \right|^2 \delta (\omega + \varepsilon_{\mu}^p - \varepsilon_{\nu}^{-p} ),
 \label{eq:spectral_inf}
\end{align}
which is an equal-weight sum over all eigenstates and where the partition function $\mathcal{Z}$ equals the Hilbert space dimension $M$.
The features  described in the previous subsection for zero temperature survive even in this extreme opposite limit, as we demonstrate explicitly by evaluating Eq.~\eqref{eq:spectral_inf} numerically.

At infinite temperature, the small energies we are interested in are in comparison to the mean level spacing
\begin{equation}
\Delta_\infty
 = \llangle \varepsilon_{\mu+1} - \varepsilon_\mu \rrangle
 = \chi \frac{\varepsilon_+ - \varepsilon_-}{M},
 \label{eq:mean_level_spacing}
\end{equation}
where $\llangle \ldots \rrangle$ denotes thermal and ensemble average,
$\chi$ is the level degeneracy, and $\varepsilon_\pm$ are the maximal and minimal energies. $\varepsilon_\pm$ can be computed analytically for all $k$:
By employing the Majorana anticommutation relations, and the mean and variance of $J_{qrst}$, Eq.~\eqref{eq:statistics_J}, we find that the variance of the energy eigenvalues
\begin{equation}
 \llangle \varepsilon_\mu^2 \rrangle
 = \frac{\mathrm{tr} \left[ \langle H^2 \rangle \right]}{M}
 = J^2 \frac{3!}{k^3} \binom{k}{4}
\end{equation}
and the mean $\llangle \varepsilon_\mu \rrangle = 0$.
Since the eigenvalues do not strictly follow a Wigner semicircle distribution~\cite{Garcia:2016il}, we need to take into account the correction~\cite{Garcia:2017dv}
\begin{equation}
\zeta = \binom{k}{4}^{-1} \sum_{r=0}^4 (-1)^r \binom{4}{r} \binom{k-4}{4-r}
\end{equation}
to obtain
\begin{equation}
\varepsilon_\pm = \pm \frac{2 \sqrt{ \llangle  \varepsilon_\mu^2 \rrangle}}{\sqrt{1-\zeta}}.
\label{eq:maximal_energy}
\end{equation}

\begin{figure}
 \includegraphics[width=\linewidth]{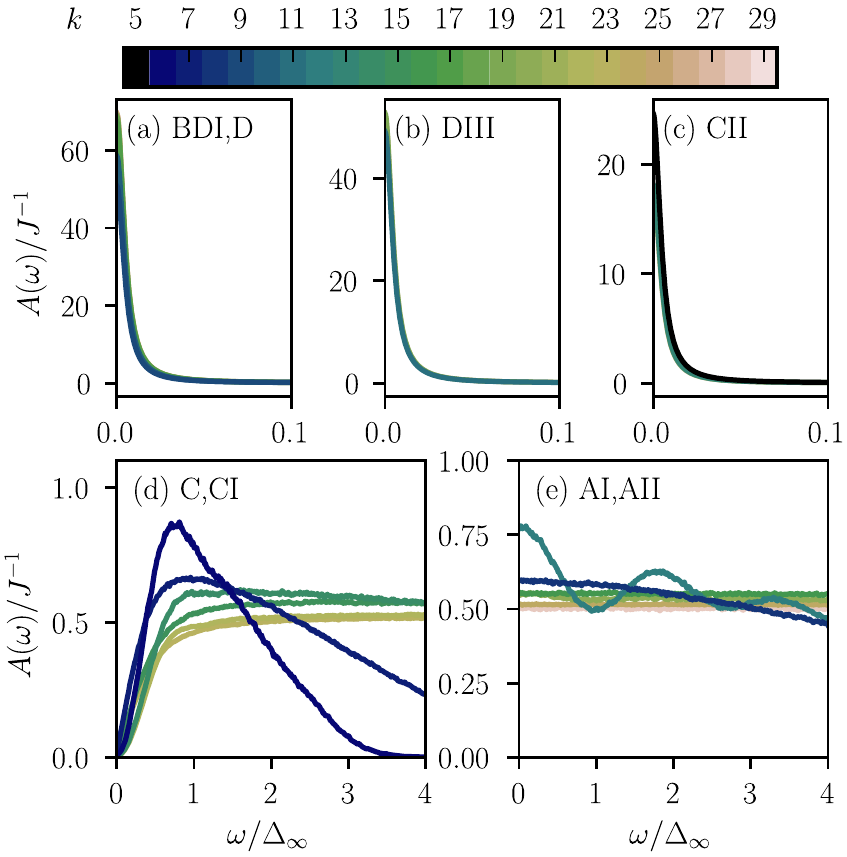}
 \caption{Spectral function at infinite temperature versus $\omega/\Delta_\infty$; cf.\ Eq.~\eqref{eq:mean_level_spacing}.
 The level broadening is $\eta = 0.005\Delta_\infty$.
 }
 \label{fig:dos_at_inf}
\end{figure}

In Fig.~\ref{fig:dos_at_inf}, we show the spectral function at infinite temperature.
The overall behavior is very similar to the spectral function at zero temperature, Fig.~\ref{fig:dos_at_zero}, i.e., we observe peaks, holes or featureless behavior at $\omega \ll \Delta_\infty$.
For large $k$, fewer realizations of $J_{qrst}$ than for zero temperature are needed to obtain smooth functions of energy (e.g., $2^{8}$ realizations for $k=28$ at $\beta^{-1} = \infty$ vs.\ $2^{11}$ at $\beta^{-1} =0$), since the number of states contributing to the spectral function close to $\omega = 0$ grows with the Hilbert space dimension at $\beta^{-1} =\infty$, while it stays constant at $\beta^{-1}=0$ (only states close to the ground states contribute).

In classes C and CI, Fig.~\ref{fig:dos_at_inf}~(d), the spectral function follows the level spacing statistics of $\varepsilon_{\mu+1} - \varepsilon_\mu$ up to energies $\omega \sim \Delta_\infty$.
As for zero energy, 
unfolding the energies reveals power laws characteristic of
the Gaussian unitary (C) and Gaussian orthogonal ensemble (CI).

As evident from panels (a)--(c), it is again the $A_q\neq 0$ cases (classes BDI, D, DIII, and CII) that display zero energy peaks. The appearance of these peaks  
may be understood by noting that the infinite temperature spectral function displays a particularly transparent relation between a $\delta(\omega)$ contribution and the zero mode piece $A_q$ of $\gamma_q$.
Specifically, we find
\begin{align}
 A (\omega ) 
 =& \frac{2}{M} \frac{1}{k} \sum_q \sum_{ \substack{p,\mu\nu \\ \varepsilon_\mu \neq \varepsilon_\nu}} \left| \bra{\psi_\mu^p} B_q \ket{\psi_\nu^{-p}} \right|^2 \delta (\omega + \varepsilon_\mu^p - \varepsilon_\nu^{-p}) \nonumber \\
  & + \frac{2}{M} \frac{1}{k} \sum_q \mathrm{tr} \left[ A_q A_q \right] \delta (\omega) .
  \label{eq:spectral_zero_mode}
\end{align}
This relation is also a useful starting point for assessing the relative weight of the $\delta(\omega)$ term compared to the nonzero energy ($\omega\gtrsim \Delta_\infty$) part of $A(\omega)$.
For $k$ sufficiently large ($k\gtrsim 10$), the ensemble average of the latter takes a roughly $\omega$ and $k$ independent value of order unity;
this translates to a weight of order $\Delta_\infty$ per eigenstate, which scales as $\propto \varepsilon_+/M$ as the function of $k$.

\begin{figure}
 \includegraphics[scale=1]{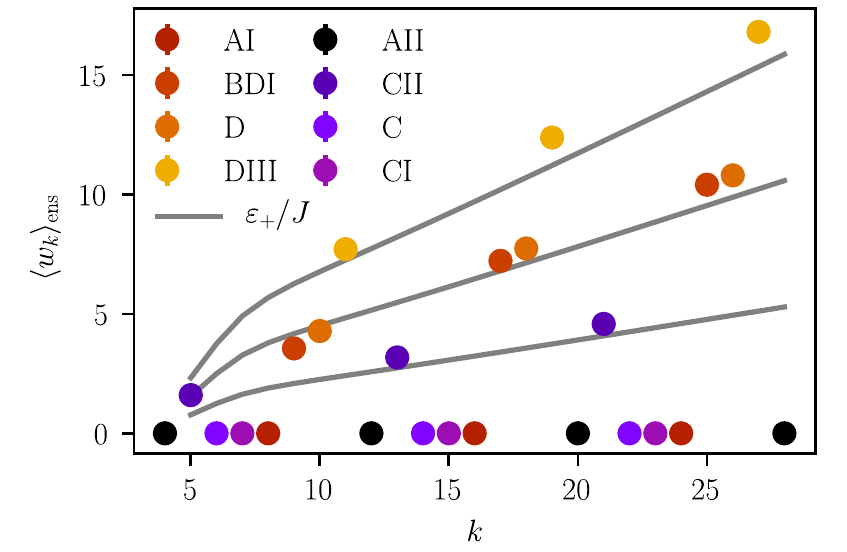}
 \caption{Ensemble-average  $\langle w_k\rangle$ as a function of $k$, where different colors denote different symmetry classes.
 For the $A_q\neq 0$ classes, $\langle w_k\rangle$ increases with $k$, approximately following $m \varepsilon_+ /J$  (solid lines included as guide for the eye) with $m=1$ for class CII, $m=2$ for classes BDI and D, and $m=3$ for class DIII.
 Error bars denoting the sample standard deviation are smaller than the marker size.
 }
 \label{fig:zero_modes}
\end{figure}

For the weight of $\delta(\omega)$ to scale identically, the ensemble average of $w_k \equiv \frac{1}{k} \sum_q \mathrm{tr} [ A_q A_q]$ must scale as \mbox{$\langle w_k\rangle \sim \varepsilon_+$} with $k$. 
As we show in Fig.~\ref{fig:zero_modes}, this is precisely the observed behavior.
For $k \gtrsim 10$, we find $\langle w_k \rangle\approx m \varepsilon_+/J$ with a simple, symmetry-class-dependent rule for the coefficient: 
$m/\chi=1$ for classes BDI and D, $m/\chi = 3/4$ for class DIII, and $m/\chi=1/4$ for class CII.

\subsection{Single-particle correlations in the time domain}

In the time domain, further signatures of the zero modes are visible in the retarded single-particle correlation function
\begin{equation}
 C^+_{qq;\rho} (t_1, t_2 ) = - i \theta(t_1 - t_2 ) \mathrm{tr} \left[ \rho \left\lbrace \gamma_q (t_1) , \gamma_q (t_2 ) \right\rbrace \right]
\end{equation}
with the density matrix $\rho$.
When $\rho$ is diagonal in the energy eigenbasis,
\begin{equation}
 \rho = \sum_{p\mu} \alpha_{p \mu} \ket{\psi_\mu^p} \bra{\psi_\mu^p},
 \label{eq:diagonal_density_matrix}
\end{equation}
e.g., $\rho = \mathcal{Z}^{-1} \exp (-\beta H)$ for a thermal average, we have $[ \rho , H ] = 0$ and hence
\begin{align}
  C^+_{qq;\rho} (t) = - i \theta(t ) \mathrm{tr} \left[ \rho \left( e^{i H t} \gamma_q e^{-i H t} \gamma_q  + (t \to -t )  \right) \right] .
 \end{align}
Now decompose $\gamma_q = A_q + B_q$ and use $[ A_q , H ] = 0$.
This results in the decomposition of the spectral function into four terms.
Cross terms that involve both $A_q$ and $B_q$ are zero,
\begin{align}
   \bra{\psi_\mu^p} & A_q \ket{\psi_\nu^{-p}} \bra{\psi_\nu^{-p}} B_q \ket{\psi_\mu^p} \nonumber \\
 &= \bra{\psi_\mu^p} B_q \ket{\psi_\nu^{-p}} \bra{\psi_\nu^{-p}} A_q \ket{\psi_\mu^p}
 = 0, 
 \label{eq:cross_terms}
\end{align}
since $\bra{\psi_{\mu}^p} A_q \ket{\psi_{\nu}^{-p}}$ is only nonzero when $\varepsilon_\nu^p = \varepsilon_{\mu}^{-p}$ while $\bra{\psi_{\nu}^{-p}} B_q \ket{\psi_{\mu}^{p}}$ is only nonzero  when $\varepsilon_\mu^p \neq \varepsilon_{\nu}^{-p}$.
Only the remaining two terms contribute to $C^+_{qq;\rho} (t )$, specifically, 
\begin{align}
  C^+_{qq;\rho} (t ) =
  &- i \theta(t) \mathrm{tr} \left[ \rho \left( e^{i H t} B_q e^{-i H t} B_q  + (t \to -t )  \right) \right] \nonumber \\
  &- 2 i \theta(t ) \mathrm{tr} \left[ \rho A_q A_q \right] .
 \end{align}
Thus, while the $B_q$ part leads to the conventional time-dependent behavior, the zero mode piece $A_q$ gives a time-independent contribution. 
For the infinite-temperature ensemble, $\rho = M^{-1}$, the time-independent part is proportional to $\mathrm{tr} [ A_q A_q ]$ that also enters in the spectral function in the energy domain, Eq.~\eqref{eq:spectral_zero_mode}.
At long times, the time-dependent contribution decays to zero, but $\mathrm{tr} \left[ \rho A_q A_q \right]$ remains: if (and only if) $A_q\neq 0$ is present, $C^+_{qq;\rho} (t )$ has a nonzero long-time limit. 

\begin{figure}
\includegraphics[scale=1]{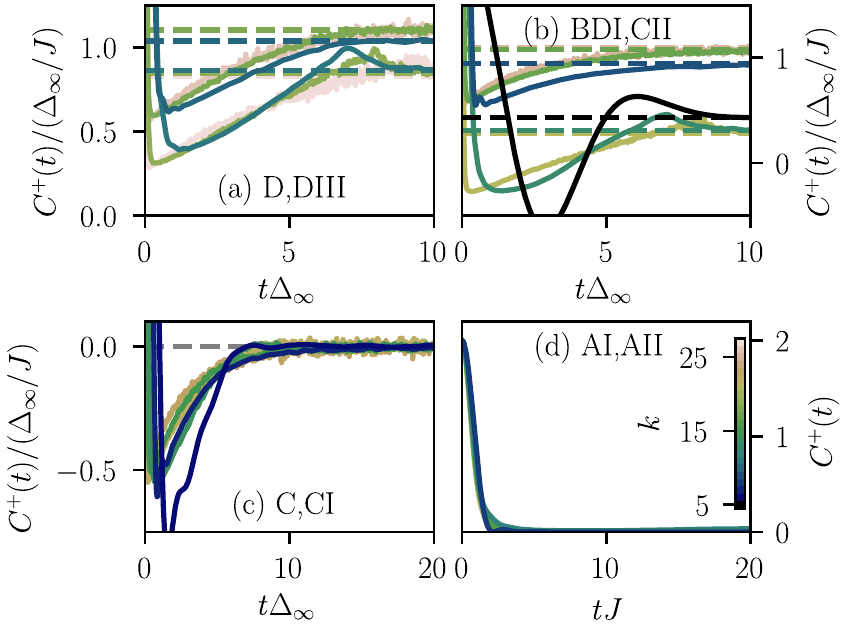}
\caption{Ensemble-averaged single-particle Green's function in the time domain, $C^+ (t)$, at infinite temperature and averaged over $2^7$ to $2^{17}$ realizations of $J_{qrst}$.
(a) In classes D and DIII, a ramp in $C^+(t)$ is visible that connects to the long-time limit with a kink (class DIII) or in a sharp corner (class D).
(b) Similarly, the connection has a kink in class CII and it is smooth in BDI.
(c) Ramps are also visible when the long-time limit equals zero, as in classes C (connection with a sharp corner) and CI (smooth connection).
(d) Only in classes AI and AII, with independent parity sectors, does the spectral function decay directly to zero, without a ramp.
Note that in panels (a)--(c) both axes are rescaled by $\Delta_\infty$.
\label{fig:time_evolution}
}
\end{figure}

Differences in the long-time limit that depend on $k$ were previously observed in Ref.~\onlinecite{Cotler:2017fx} (for even $k$) and linked to equal-energy matrix elements.
As additionally pointed out there, degenerate parity sectors give ``ramps'' in the time-domain, i.e., the correlations linearly increase to their long-time limit after an initial dip~\cite{Cotler:2017fx,Guhr:1998bg}.
Such ramps are visible in all symmetry classes with degenerate parity sectors, i.e., they are only missing in classes AI and AII.
The ramps connect differently to the long-time limit, depending on the level spacing statistics:
either with a kink (GSE, classes DIII and CII), smoothly (GOE, classes BDI, CI), or a sharp corner (GUE, classes D and C)~\cite{Guhr:1998bg}.
Our findings show that both long-time limits and ramps go hand in hand with the zero modes:
the presence of odd-parity zero modes corresponds to ramps, and $A_q \neq 0$ corresponds to nonzero long-time limits.
We illustrate this in Fig.~\ref{fig:time_evolution}, where we show $C^+ (t) \equiv i \frac{1}{k} \sum_q C^+_{qq} (t)$ at infinite temperature and averaged over a large ensemble, thus linking the behavior of the single-particle Green's function in the time domain to the eight (real) Altland-Zirnbauer classes.

\subsection{Out-of-time-order correlation function}

How do the zero modes influence other, including multiparticle, correlation functions?
In this section, we focus on the out-of-time-order correlation function and investigate how the low-energy features expected to arise due to zero modes translate into the long-time behavior.

We consider the OTOC that is symmetrized with respect to the thermal weights~\cite{Bagrets:2017gk,Maldacena:2016gp}
\begin{align}
 F_{qr} (t) = \mathcal{Z}^{-1} \mathrm{tr} & \left[ \gamma_q y \gamma_r y^\dagger \gamma_q y \gamma_r y^\dagger \right]
 \label{eq:otoc_general}
\end{align}
where the operator
\begin{equation}
 y \equiv \exp \left[ - \left(i t + \tfrac{1}{4}\beta \right) H \right]
\end{equation}
includes both real and imaginary-time evolution.
We again split up the Majorana operators into zero-mode and nonzero-energy contributions, $\gamma_q = A_q + B_q$, and use $\left[ A_q, y \right] = 0$ to rewrite the terms contributing to the trace of Eq.~\eqref{eq:otoc_general}, e.g.,
\begin{equation}
 \mathrm{tr} \left[ A_q y A_r y^\dagger A_q y A_r y^\dagger \right] =
 \mathrm{tr} \left[ A_q A_r A_q A_r y y^\dagger y y^\dagger \right]
\end{equation}
with the time-independent thermal weight $y y^\dagger = \exp ( - \beta H /2)$.
Similarly, the contributions containing two $A_q$ or two $A_r$ operators are time-independent,
\begin{align}
 \mathrm{tr} \left[ A_q y B_r y^\dagger A_q y B_r y^\dagger \right] &=
 \mathrm{tr} \left[ A_q B_r y y^\dagger A_q B_r y y^\dagger \right] \\
 \mathrm{tr} \left[ B_q y A_r y^\dagger B_q y A_r y^\dagger \right] &=
 \mathrm{tr} \left[ B_q A_r y y^\dagger B_q A_r y y^\dagger \right] .
\end{align}
All other terms are either zero, e.g.,
\begin{equation}
 \mathrm{tr} \left[ A_q y A_r y^\dagger A_q y B_r y^\dagger \right] =
 \mathrm{tr} \left[ A_q A_r A_q B_r y^\dagger y y^\dagger y \right] = 0,
\end{equation}
as can be seen by inserting a complete basis [cf.\ Eq.~\eqref{eq:cross_terms}], or time dependent, since $[B_q,y] \neq 0$, such that terms $B_q y B_r$ are always time dependent.
Hence, the OTOC also decomposes into a time-independent and a time-dependent part,
\begin{equation}
  F_{qr} (t) = F_{qr}^{\infty} + \delta F_{qr} (t),
\end{equation}
with
\begin{align}
\label{eq:otoc_decomposition}
 F_{qr}^{\infty}
 &= \mathcal{Z}^{-1} \left( \mathrm{tr} \left[ A_q A_r A_q A_r y y^\dagger y y^\dagger \right] \right.  \\
 & \left. + \mathrm{tr} \left[ A_q B_r y^\dagger y A_q B_r y^\dagger y \right] + \mathrm{tr} \left[ B_q A_r y y^\dagger B_q A_r y y^\dagger \right] \right). \nonumber
\end{align}
We emphasize that $F_{qr}^{\infty}\neq 0$ if and only if $A_q\neq 0$.

\begin{figure}
 \includegraphics[scale=1]{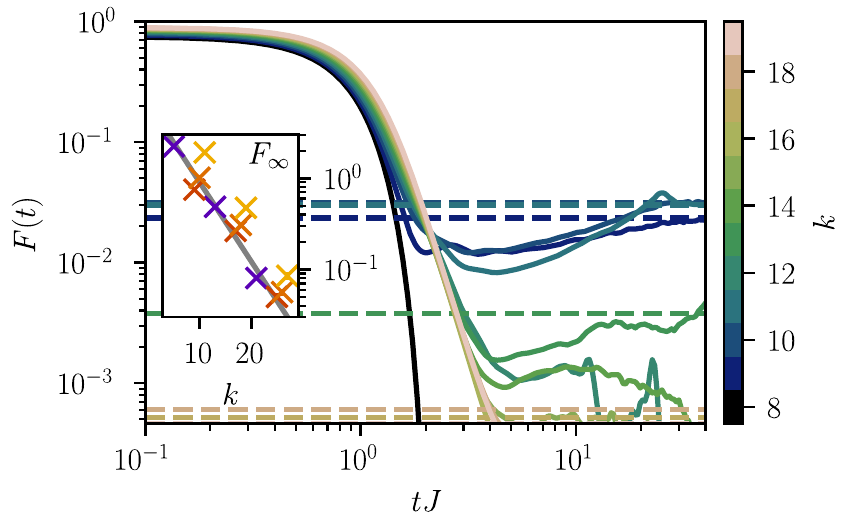}
 \caption{Ensemble-averaged OTOC $F(t)$ as a function of time for different $k$.
 When $A_q = 0$, the OTOC approaches zero for long times, here clearly visible for $k=8$.
 Otherwise, after some initial decay, the OTOC approaches a long-time value $F_\infty/M$, marked by the dashed lines.
 We chose a log-log scale for a better visibility of the different time scales.
 The inset shows $F_\infty$ as a function of $k$, with the gray line an exponential decay as a guide for the eyes.
 Different colors denote different symmetry classes.
 }
 \label{fig:otoc}
\end{figure}

In Fig.~\ref{fig:otoc}, we show the numerically obtained OTOC at infinite temperature, averaged over a large ensemble ($2^{7}$ to $2^{16}$ realizations of $J_{qrst}$) and all Majorana modes, 
\begin{equation}
 F(t) = \frac{1}{k^2} \frac{1}{M} \sum_{q,r} \langle \mathrm{tr} \left[ \gamma_q  (t) \gamma_r \gamma_q (t) \gamma_r \right] \rangle,
 \label{eq:Finfty}
\end{equation}
with $\gamma_q (t) = \exp ( i t H) \gamma_q \exp (- i t H)$.
While for classes with $A_q=0$ the absence of $F_{qr}^\infty$  implies that $F(t)$ decays to zero, 
for all classes with $A_q \neq 0$ the consequent \mbox{$F_{qr}^\infty \neq 0$} leads to a nonzero long-time limit for $F(t)$ (dashed lines in Fig.~\ref{fig:otoc}).
In the inset, we show the scaling of $F_\infty  \equiv M \lim\limits_{t \to \infty} F(t)$ with $k$.
The numerical data suggest an approximately exponential decay of $F_\infty$, although we cannot determine the exact scaling due to the finite system sizes within numerical reach.

\section{Symmetry-breaking terms}
\label{sec:bilinear}

How do the properties shown above survive once symmetry-breaking terms are added to the Hamiltonian?
We investigate an additional random bilinear term~\cite{Maldacena:2016hu,Chen:2017fc,Song:2017ka,Garcia:2018ck,Eberlein:2017jb} that extends the original model~\eqref{eq:hamiltonian_SYK}
\begin{equation}
 H = i \sum_{q<r} K_{qr} \gamma_q \gamma_r + \sum_{q<r<s<t} J_{qrst} \gamma_q \gamma_r \gamma_s \gamma_t
 \label{eq:hamiltonian_KJ}
\end{equation}
and defines a second energy scale $K$ via
\begin{align}
 \left\langle K_{I} \right\rangle = 0 , & &
 \left\langle K_I K_{I'} \right\rangle = \frac{1}{k} K .
 \label{eq:statistics_K}
\end{align}
This bilinear term anticommutes with both $T_+$ and $T_-$, i.e., it breaks both antiunitary symmetries.
When viewing our system as the end of a one-dimensional noninteracting system, this corresponds to breaking time-reversal symmetry in the one-dimensional bulk~\cite{Fidkowski:2010ko,Fidkowski:2011dh}, changing its symmetry class from BDI to D.	
In the noninteracting case, this allows two distinct topological sectors characterized by a $\mathbb{Z}_2$ invariant~\cite{Kitaev:2000gb,Schnyder:2008ez}.
Once we introduce $K_{qr}$, pairs of Majorana modes gap out, leaving either one or no zero-energy mode, hence, the $\mathbb{Z}_2$ classification.
In the presence of interactions, the $\mathbb{Z}_2$ classification remains~\cite{Gangadharaiah:2011bt,Katsura:2015ep}.
Using our approach with the $T_+$ and $T_-$ symmetries, this $\mathbb{Z}_2$ classification can be also given a different interpretation:
even though both antiunitary symmetries are broken, the product of the two, the chiral symmetry $Z=T_+ T_-$, remains for odd $k$.
This allows one to view the odd $k$ case as implementing the chiral unitary symmetry class AIII, and the case with $k$ even, with no symmetries, the unitary class A.
The latter case was discussed in Ref.~\onlinecite{Kanazawa:2017be}.

In class AIII, the Majorana operators can again be split up into a zero-mode and remainder contribution, $\gamma_q = A_q + B_q$, due to the guaranteed parity degeneracy.
Since $Z$ changes the parity of a state, the combination $\gamma_q Z$ preserves parity.
Thus, the matrix element
\begin{equation}
 \bra{\psi^p_\mu} \gamma_q Z \ket{\psi^p_\mu} \neq 0 ,
 \label{eq:peak_AIII}
\end{equation}
which implies that a peak at zero energy is present in the spectral function for odd $k$ even when $K \neq 0$.

\begin{figure}
 \includegraphics[width=\linewidth]{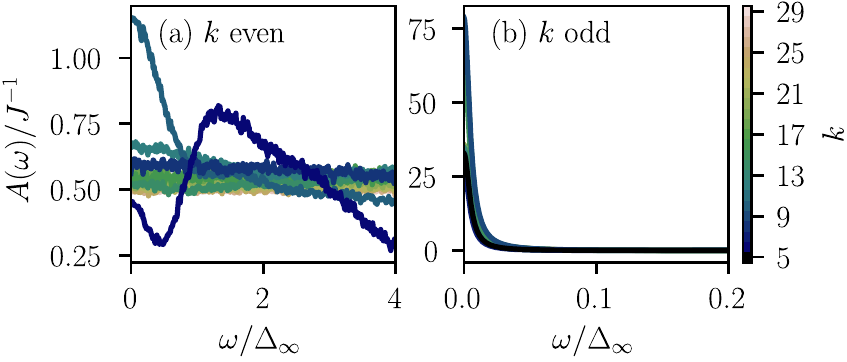}
 \caption{Spectral function at infinite temperature for the SYK model including bilinear terms~\eqref{eq:hamiltonian_KJ} with $J=K$; cf. Eqs.~\eqref{eq:statistics_J} and~\eqref{eq:statistics_K}.
 (a) For even $k$, the system is in symmetry class A.
 The spectral function is almost featureless (oscillations die out with increasing $k$).
 (b) For odd $k$, the chiral symmetry $Z$ remains and the system is in symmetry class AIII.
 The spectral function displays a peak at zero energy; cf.\ Eq.~\eqref{eq:peak_AIII}.
 }
 \label{fig:dos_with_K}
\end{figure}

In class A, on the other hand, the different parity sectors are nondegenerate.
The operator $Z$ does not exist and the overlap of equal-energy states always vanishes,
\begin{equation}
 \bra{\psi^p_\mu} \gamma_q \ket{\psi^p_\mu} = 0 ,
\end{equation}
giving a featureless spectral function close to zero energy.

In Fig.~\ref{fig:dos_with_K}~(a), we show that the spectral function~\eqref{eq:spectral_function} at $\omega \sim 0$ is indeed featureless for even $k$, class A.
As expected from the previous considerations, the spectral function has a peak for odd $k$, class AIII, shown in panel~(b).
Similarly, the OTOC (not shown) does not decay to zero in class AIII, while it does in class A.
It can be decomposed into a time-dependent and a time-independent part, with the latter given by Eq.~\eqref{eq:otoc_decomposition}.

In addition to the eight classes discussed in the absence of $K$, we thus see that the physics that arises by adding $K$ (or indeed by breaking $T_+$ and $T_-$ in different ways, e.g., by six-Majorana terms) may be viewed as bringing about the two complementary complex symmetry classes, thereby completing the tenfold Altland-Zirnbauer classification.

\section{Conclusion and outlook}
\label{sec:conclusion}

In this work we brought together the SYK model and a corresponding many-body incarnation of the Altland-Zirnbauer classification. 
We discovered that the SYK model furnishes many-body counterparts of features familiar from Altland-Zirnbauer classes of the single-particle domain, and that these have interesting consequences for the strongly-correlated dynamics of the system.
In particular, we found that the Altland-Zirnbauer symmetry classes manifest themselves not only in the level spacing statistics (that distinguishes them only along the three Wigner Dyson classes~\cite{Garcia:2016il,You:2017jj,Cotler:2017fx}), but also in the presence or absence of zero modes and their single-particle weight $A_q$ in Majorana operators $\gamma_q$.
These have been found to set the dominant features of a range of correlation functions at low energies (or long times).

The single-particle spectral function has been found to display a peak (hole) at zero energy whenever the SYK model supports zero modes with nonzero (zero)  $A_q$.
The zero modes also determine the behavior at small nonzero energies:
they lead to power-law scaling in the spectral-function with the exponent determined by the level repulsion (as set by time-reversal $T_+$), corresponding to the system's symmetry class.
All of these features are absent in the cases without zero modes.
These symmetry class dependent characteristics are in a one-to-one correspondence to analogous Altland-Zirnbauer features of the density of states in the free-fermionic case.
We emphasize  that this is \textit{not} due to Altland-Zirnbauer features of some free fermion system surviving the addition of interactions, but rather a remarkable pattern of signatures emerging in an interaction-only model.

The effects on the strongly correlated dynamics have been illustrated by considering correlation functions in the time-domain, including the out-of-time-order correlation function (OTOC).
We found that, while the OTOC displays its familiar long-time decay to zero for classes where zero modes are absent or have vanishing single-particle weight, the presence of zero modes with $A_q\neq 0$ dictates that the OTOC reaches a nonzero long-time limit.
As these OTOC results illustrate, the main dynamical effect of zero modes is on the long-time behavior.
We expect zero modes to exhibit interesting long-time effects also on other dynamical correlation functions, the study of which we leave for future work.

The features we have summarized so far have been for a purely quartic SYK model, thus in the presence of the antiunitary symmetry $T \gamma_q T^{-1}=\gamma_q$;
depending on the number $k$ of Majoranas, these systems realize the eight real Altland-Zirnbauer classes.
We have shown, however, that the remaining two complex Altland-Zirnbauer classes also arise in the SYK model, if one  allows terms (e.g., quadratic couplings) that explicitly break $T$.
In this case, for even $k$, all symmetries are broken and the system is in class A, while for odd $k$, chiral symmetry remains and the system is effectively in class AIII.
In the latter case, the two parity sectors remain degenerate, which results in an odd-parity zero mode that has a nonzero overlap with the Majorana operators.
Consequently, the spectral function in class AIII displays a peak at zero energy, whereas it is featureless in class A.
Thus, allowing for both $T$ preserving (e.g., quartic) and $T$-breaking (e.g. quadratic, or six-Majorana) terms, the SYK model realizes the entire tenfold way.

We emphasize that the zero modes we find are not low-energy objects:
they have consequences for all states, not only the ground state~\cite{Fendley:2016ie}.
We have demonstrated this by evaluating correlation functions with respect to all states, in an equal weight average, i.e. at infinite temperature:
we found that both the single-particle spectral function and the OTOC display their aforementioned features, even in this extreme limit.
Furthermore the correlation functions need not even be thermal:
we expect analogous low-energy (long-time) features to arise in the highly nonequilibrium situation where the reference state for the correlation functions is an (arbitrary) highly excited state $|\psi_\mu^p\rangle$, replacing the groundstate $|\psi^p_0\rangle$, e.g., in Eq.~\eqref{eq:spectral_function_T0}.

The zero modes we find in fact are ``strong''~\cite{Fendley:2016ie} (even exact) zero modes within the SYK model.
However, if one views the SYK  model as arising in a higher dimensional system (e.g., at the end of a one-dimensional system or in a vortex), they are not strong zero modes any more, as they arise within the  low-energy physics of this larger system.
While they are thus low-energy objects from this perspective, they may provide useful starting points for constructing strong zero modes in these cases. 

Of the signatures we have described, the low-energy behavior of the spectral function is the most accessible in experiments.
In the proposed realizations of the SYK model in the solid-state~\cite{Pikulin:2017js,Chew:2017fn,Chen:2018ho}, the spectral function may be directly measured in tunneling setups consisting of an electrode weakly coupled to the Majorana fermions~\cite{Pikulin:2017js,Chew:2017fn,Chen:2018ho,Pikulin:2015ea}.
The main limitation in this case is due to the thermally smeared variant of the spectral function appearing in the differential conductance~\cite{BruusBook};
thus, while the features we predict are robust even for infinite temperatures, to resolve the spectral function near $\omega=0$ in tunneling setups, the temperature should be of the order of the many-body level spacing or below.

Our work opens several interesting directions for future study and we close by mentioning just a few examples of these.
An immediate generalization of our work is to the recently discussed variants of the SYK model with terms coupling an even number $q>4$ Majoranas~\cite{Maldacena:2016hu,Garcia:2017dv}.
Although the symmetry classification remains valid for general $q$ (eight real classes if only terms $q=4,8, \ldots$ are included, two complex classes otherwise), the resulting observables, e.g., the scaling of the weight of the zero mode peaks relative to their background, may change.

Our results can also be generalized to systems with a local Hamiltonian instead of the all-to-all couplings in the SYK model.
For translationally invariant one-dimensional systems, bilinear terms coupling neighboring sites correspond to the \citeauthor{Kitaev:2000gb} chain and can be transformed into the transverse field Ising model~\cite{Kitaev:2000gb}.
Similarly, quartic couplings translate to an exotic spin model with nearest and next-nearest neighbor interactions~\cite{Liu:2017ea,Liu:2017bd,Monthus:2018ih}.
All results we found are also valid for these spin systems, where the local structure of the interactions may have potentially interesting consequences, e.g., an emergent nontrivial spatial structure for the zero modes.

One may also develop approaches analogous to ours for the so-called complex SYK model~\cite{Sachdev:1993hv,You:2017jj,Iyoda:2018bv}.
This model has ordinary (i.e., complex), instead of Majorana (real) fermions, and may be viewed as arising at the end of a one-dimensional system that admits a $\mathbb{Z}_4$ topological classification.
The complex SYK model shows a corresponding fourfold pattern for the three possible Wigner-Dyson level spacing statistics~\cite{You:2017jj}; 
it is an interesting question whether these cases may be further distinguished in terms of zero modes and correlation functions.

\begin{acknowledgments}
We thank Paul Fendley, Tal\'{i}a L.~M.~Lezama, and Dmitry Pikulin for useful discussions.
This work was supported by the ERC Starting Grants No. 679722 QUANTMATT and No. 678795 TopInSy, the Knut and Alice Wallenberg Foundation 2013-0093, the National Science Foundation under Grant No. NSF PHY-1748958, and the Royal Society.
\end{acknowledgments}

\appendix

\section{Explicit construction of time-reversal and particle-hole symmetry}
\label{sec:symmetries}

\begin{table}
 \begin{tabular}{c|c|c|c|c|c|c}
 \toprule
  $k \mod 8$ & ~$C_+ C_+^*$~ & ~$C_- C_-^*$~ & ~$a$~ & ~$T_+^2$~ & ~$T_-^2$~ & ~Cartan label \\
  \colrule
  0	& $+1$ & $+1$ & $+1$ & $+1$ & $0$  & AI \\
  1	& $+1$ & $0$  & $+1$ & $+1$ & $+1$ & BDI \\
  2	& $+1$ & $-1$ & $-1$ & $0$  & $+1$ & D \\
  3	& $0$  & $-1$ & $-1$ & $-1$ & $+1$ & DIII \\
  4	& $-1$ & $-1$ & $+1$ & $-1$ & $0$  & AII \\
  5	& $-1$ & $0$  & $+1$ & $-1$ & $-1$ & CII \\
  6	& $-1$ & $+1$ & $-1$ & $0$  & $-1$ & C \\
  7	& $0$  & $+1$ & $-1$ & $+1$ & $-1$ & CI\\
  \botrule
 \end{tabular}
 \caption{Eightfold periodicity of the operators $C_\pm$ and $T_\pm$.
 Depending on the number of interacting Majorana modes $k$, the matrices $C_\pm$ may not exist (denoted by $0$), or exist and square to $+1$ or $-1$.
 From $C_+$ may commute ($a=+1$) or anticommute ($a=-1$) with the parity operator $P$, such that the antiunitary operator $C_+\mathcal{K}$ effectively acts as either time-reversal ($T_+$ for $a=+1$) or particle-hole symmetry ($T_-$ for $a=-1$).
 }
 \label{tab:cmatrices}
\end{table}

In this Appendix, we show how the operators for time-reversal symmetry $T_+$ and particle-hole symmetry $T_-$ with $[ T_\pm ,\gamma_q ] = 0$ and $T_\pm P T_\pm^{-1} = \pm P$ can be constructed explicitly.
Since Majorana operators realize a Clifford algebra with signature $(+ + + + \ldots )$, they can be represented by higher-dimensional gamma matrices of dimension $M \times M$, where $M= 2^{\lfloor k/2 \rfloor}$ with $\lfloor \cdots \rfloor$ the floor function~\cite{de1986field}.
We use a basis where complex conjugation $\mathcal{K}$ does not affect Majorana modes with an even index $\mathcal{K} \gamma_{2q} \mathcal{K} = \gamma_{2q}$, but changes the sign of Majorana modes with an odd index $\mathcal{K} \gamma_{2q+1} \mathcal{K} = -\gamma_{2q+1}$.
Summarizing previous work on higher-dimensional Clifford algebras~\cite{de1986field,Brauer:1935ji,Pais:1962is,Kennedy:1981eu}, we discuss the properties of the Hilbert space spanned by $k$ Majorana operators and link these properties to the Altland-Zirnbauer classification~\cite{Altland:1997cg,Fidkowski:2011dh}
We further connect to the notation of Ref.~\onlinecite{Fidkowski:2011dh}.

\subsection{Even $k$}

For an even number of Majoranas, one can find two unitary matrices $C_+$ and $C_-$ with the properties~\cite{de1986field,Pais:1962is}
\begin{equation}
  C_\pm\,\gamma_q^* \,C_\pm^{\dagger} = \pm \gamma_q .
 \label{eq:cplusminus}
\end{equation}
As the Hamiltonian only includes parity-conserving terms involving an even number of Majoranas, the combination of each of these matrices with complex conjugation commutes with the Hamiltonian $\left[ C_\pm \mathcal{K}, H \right] = 0 $.
We can construct both matrices $C_\pm$ as the product of $k/2$ Majorana operators.
More explicitly, the form of $C_+$ depends on $k$,
\begin{equation}
 C_+ = \begin{cases} \gamma_0 \gamma_2 \cdots \gamma_{k-2} & k = 4 n + 2 \\ \gamma_1 \gamma_3 \cdots \gamma_{k-1} & k = 4 n . \end{cases}
 \label{eq:cplus}
\end{equation}
When $k=4n+2$, $C_+$ contains an odd number of Majorana operators with an even index, giving a sign change $C_+ \gamma_{2q + 1} C_+^\dagger = - \gamma_{2q + 1}$ while the sign of $\gamma_{2q}$ remains invariant as $C_+$ contains $\gamma_{2q}$ itself.
Similarly, when $k=4n$, $C_+$ contains an even number of Majorana operators with an odd index, giving a sign change $C_+ \gamma_{2q + 1} C_+^\dagger = - \gamma_{2q + 1}$ (as $C_+$ contains $\gamma_{2q+1}$ itself), while the sign of $\gamma_{2q}$ remains invariant.
Analogously,$C_-$ is given by $k/2$ Majorana operators
\begin{equation}
 C_- = \begin{cases} \gamma_1 \gamma_3 \cdots \gamma_{k-1} & k = 4 n + 2 \\ \gamma_0 \gamma_2 \cdots \gamma_{k-2} & k = 4 n . \end{cases}
 \label{eq:cminus}
\end{equation}
As both $C_+$ and $C_-$ contain $k/2$ Majorana operators, these matrices are purely real when $k=4n$, which results in
\begin{align}
 C_+ C_+^*
 &= \gamma_1 \gamma_3 \cdots \gamma_{k-1} \gamma_1 \gamma_3 \cdots \gamma_{k-1} \nonumber \\
 &= (-1)^{k/4 (k/2 -1)} = (-1)^{n} \\
 C_- C_-^*
 &= \gamma_0 \gamma_2 \cdots \gamma_{k-2} \gamma_0 \gamma_2 \cdots \gamma_{k-2} \nonumber \\
 &= (-1)^{k/4 (k/2 -1)} = (-1)^{n}
\end{align}
where the sign results from $(k/2-1 ) + (k/2-2) +  \cdots + 1 = k/4 (k/2-1)$ exchanges that are necessary to couple all Majoranas $\gamma_q \gamma_q = 1$.
Thus, when $k = 8m$ ($m\in \mathbb{N}$, $n$ even), $C_\pm C_\pm^* = 1$, while, when $k= 8m + 4$ ($n$ odd), $C_\pm C_\pm^* = -1$.
When $k = 4n +2$, $C_+$, which only contains Majoranas with an even index, is purely real, while $C_-$, which contains an odd number of Majoranas with an odd index, is purely imaginary, resulting in
\begin{align}
 C_+ C_+^*
 &= \gamma_0 \gamma_2 \cdots \gamma_{k-2} \gamma_0 \gamma_2 \cdots \gamma_{k-2} \nonumber \\
 &= (-1)^{k/4 (k/2 -1)} = (-1)^{n} \\
 C_- C_-^*
 &=-\gamma_1 \gamma_3 \cdots \gamma_{k-1} \gamma_1 \gamma_3 \cdots \gamma_{k-1} \nonumber \\
 &=-(-1)^{k/4 (k/2 -1)} =-(-1)^{n}
\end{align}
such that, when $k=8m + 2$, $C_+ C_+^* = +1$ and $C_- C_-^* = -1$, while, when $k=8m + 6$, $C_+ C_+^* = -1$ and $C_- C_-^* = +1$.
The squares of the operators are summarized in the first three columns of Table~\ref{tab:cmatrices}.

Combining both signs from Eq.~\eqref{eq:cplusminus} gives
\begin{equation}
 C_+ C_- \gamma_q C_-^{\dagger} C_+^\dagger = -C_+ \gamma_q^* C_+^\dagger = -\gamma_q ,
\end{equation}
i.e., the joint operator $C_+ C_-$ anticommutes with all $\gamma_q$.
Since the operator $\gamma_q$ changes the parity $p$ of a state $\ket{\psi^p}$~\cite{Kitaev:2000gb}, $C_+ C_-$ equals the parity operator up to a phase, $P = C_+ C_- e^{i\phi}$.
The phase is chosen such that the parity operator itself is Hermitian, which is satisfied by
\begin{equation}
 P = i^{k/2} \gamma_0\gamma_1 \cdots \gamma_{k-1} ,
\end{equation}
it equals the product of all Majorana operators times a phase~\cite{de1986field}.
Given the definition of $P$, we realize that, since $C_+ \mathcal{K}$ commutes with all $\gamma_q$, only the number of Majorana operators $k$ determines if $C_+ \mathcal{K}$ and $P$ commute or anticommute
\begin{equation}
 C_+ P^* C_+^\dagger = (-1)^{k/2} P \equiv a P .
 \label{eq:parity_tr}
\end{equation}
Thus, for $k = 4n$, $C_+ \mathcal{K}$ and $P$ commute, such that time-reversal symmetry is present with $T_+ = C_+ \mathcal{K}$.
For $k=4n + 2$, $C_+ \mathcal{K}$ and $P$ anticommute, such that particle-hole symmetry is present with $T_- = C_+ \mathcal{K}$.
Using that $P$ equals $C_+ C_-$ up to a phase, it is evident that $a$ equals the product of $(C_- \mathcal{K})^2$ and $(C_+ \mathcal{K})^2$
\begin{equation}
 a = C_+ P^* C_+^\dagger \hat{P} = 
 C_+ C_+^* \, C_-^* C_-,
\end{equation}
with $a$ given in Table~\ref{tab:cmatrices} for different numbers of Majoranas $k$.

In Ref.~\onlinecite{Fidkowski:2011dh}, both $T_-$ and $T_+$ are referred to as time-reversal symmetry and denoted by $\hat{T}$.
This notation stems from the system that is considered, i.e., the edge of a one-dimensional BDI system.
The \emph{physical} symmetry that restricts the interaction is time-reversal symmetry---bilinear terms, such as in Eq.~\eqref{eq:hamiltonian_KJ}, break $\hat{T}$, but not the physical particle-hole symmetry.
When only considering the edge, $\hat{T}$ breaks down to either particle-hole or time-reversal symmetry.

\subsection{Odd $k$}

As discussed in the main text, we need to include an additional Majorana fermion located at infinity that is completely decoupled from the Hamiltonian.
Parity is only well-defined when $\gamma_\infty$ is included, giving
\begin{equation}
 P = i^{(k+1)/2} \gamma_0 \gamma_1 \cdots \gamma_{k-1}\gamma_\infty .
\end{equation}
Chiral symmetry is represented by the operator
\begin{equation}
 Z = i^{(k-1)/2} \gamma_0 \gamma_1 \cdots \gamma_{k-1}
\end{equation}
that anticommutes with $P$ and commutes with all $\gamma_q$.

Different from even $k$, only one of the matrices $C_+$ and $C_-$ can be found without incorporating $\gamma_\infty$.
This can be easily seen from considering both operators for $l=k+1$ Majorana modes, Eqs.~\eqref{eq:cplus} and~\eqref{eq:cminus}:
the operator $\gamma_\infty \equiv \gamma_{l-1}$ is necessary to construct $C_+$ for $l=4n$ and $C_-$ for $l=4n+2$.
Thus, only $C_+$ can be found when $k=4 n + 3$ and only $C_-$ when $k=4n+1$.

Incorporating $\gamma_\infty$, however, gives the same operators $C_\pm$ as for $l=k+1$ Majoranas.
Depending on $k$, the operators $C_\pm \mathcal{K}$ and $Z$ may commute or anticommute,
\begin{equation}
 C_\pm Z^* C_\pm = \pm (-1)^{(k-1)/2} Z \equiv \pm a Z
\end{equation}
with the sign given by the phase of $Z$ plus an extra minus sign $(-1)^k$ for $C_- \mathcal{K}$.
As $\gamma_\infty$ is not part of the local Hilbert space, we can define the action of $C_\pm$ on $\gamma_\infty$ most conveniently; following Ref.~\onlinecite{Fidkowski:2011dh}, we chose
\begin{equation}
 C_+ \gamma_\infty^* C_+^\dagger = - a \gamma_\infty .
\end{equation}
This choice ensures $[ C_+ \mathcal{K}, P] = 0$, such that $T_+ = C_+ \mathcal{K}$ for all odd $k$.
In Ref.~\onlinecite{Fidkowski:2011dh}, $T_+$ is identified with the physical time-reversal $\hat{T}$.
Different from even $k$, the Hamiltonian commutes with $Z$, i.e., we can construct a second antiunitary operator $T_- = T_+^{-1} Z$.
Since $\lbrace Z,P \rbrace = 0$, $\lbrace T_-,P \rbrace = 0$ is always guaranteed.
The square of $T_+$ is identical to the square $C_+C_+^*$ for $l=k+1$ Majoranas.
The square of $T_-$ is
\begin{equation}
 T_-^2 = T_+^{-1} Z T_+^{-1} Z = T_+^2 T_+ Z T_+^{-1} Z = a T_+^2 .
\end{equation}
The operators $C_\pm$, $T_\pm$ and the sign of the commutation relation are summarized in Table~\ref{tab:cmatrices}.

\bibliography{syk}

\end{document}